# Gamify Employee Collaboration - A Critical Review of Gamification Elements in Social Software
## (*Research-in-Progress*)

**Christian Meske**
Department of Computer Science and Applied Cognitive Science
University of Duisburg-Essen
Duisburg, Germany
Email: christian.meske@uni-due.de

**Tobias Brockmann**
Department of Computer Science and Applied Cognitive Science
University of Duisburg-Essen
Duisburg, Germany
Email: tobias.brockmann@uni-due.de

**Konstantin L. Wilms**
Department of Computer Science and Applied Cognitive Science
University of Duisburg-Essen
Duisburg, Germany
Email: konstantin.wilms@uni-due.de

**Stefan Stieglitz**
Department of Computer Science and Applied Cognitive Science
University of Duisburg-Essen
Duisburg, Germany
Email: stefan.stieglitz@uni-due.de

## Abstract

Social software solutions in enterprises such as IBM Connections are said to have the potential to support communication and collaboration among employees. However, companies are faced to manage the adoption of such collaborative tools and therefore need to raise the employees' acceptance and motivation. To solve these problems, developers started to implement Gamification elements in social software tools, which aim to increase users' motivation. In this research-in-progress paper, we give first insights and critically examine the current market of leading social software solutions to find out which Gamification approaches are implemented in such collaborative tools. Our findings show, that most of the major social collaboration solutions do not offer Gamification features by default, but leave the integration to a various number of third party plug-in vendors. Furthermore we identify a trend in which Gamification solutions majorly focus on rewarding quantitative improvement of work activities, neglecting qualitative performance. Subsequently, current solutions do not match recent findings in research and ignore risks that can lower the employees' motivation and work performance in the long run.

## 1   Introduction

During the last decade social media massively invaded our daily lives and equally affected the business world (Larosiliere et al. 2015; Kane et al. 2014). Facebook, Twitter, LinkedIn, Google+, and YouTube are used for marketing activities and professional communication of organizations, political actors and celebrities (Chui et al. 2012; Kane et al. 2014; Stieglitz et al. 2014a). Besides this external scope, social media principles and functions are increasingly used inside organizations to support communication and collaboration among employees (Backhouse 2009). Thereto social media encourage, mainly through information and knowledge sharing, the support of communication, workflows and creative groups (Chui et al. 2012; Meske et al. 2014). Software supporting this purpose is often called social collaboration tools, (enterprise) social software or enterprise social media. Organizations are faced to manage the adoption of such social collaboration tools and need to promote social software and find mechanisms to raise the employees' acceptance and usage (Meske and Stieglitz 2013; Stieglitz and Meske 2012). Moreover, Pawlowski et al. (2014) identified in a structured literature review that the technical adoption and acceptance of technologies is one of the major issues using social software.

One approach to improve the employees' acceptance, motivation and usage of software is Gamification (Deterding et al. 2011). This approach is frequently examined in different professionals fields (e.g. innovation management, knowledge sharing) from a multitude of researchers (Kaleta et al. 2014; Lounis et al. 2014; Teh et al. 2013). Gamification describes the application of typical game elements like high scores, badges, or virtual goods to traditional non-game contexts (e.g. learning, work). According to Stieglitz (2015) 'Enterprise Gamification' is defined as the integration of playful elements into business process or into the learning environment of enterprises. Gamification works, as it makes use of basic human needs (e.g. success, reward, status, competition, self-expression, altruism) (Thiebes et al. 2014). Hence Gamification helps to raise the extrinsic motivation of employees. However, although Gamification has proven to be effective in the context of various Information Systems, this approach has also been criticized for diminishing users' motivation. For example Amriani et al. (2013) showed that elements like points, badges and leaderboards could diminish the intrinsic motivation of the users, since these elements only support extrinsic motivation. This effect of secondary extrinsic motivation lowering the primary intrinsic motivation is known as the "overjustification effect" (DeCharms 1968).

Enterprises, often start-ups, began to develop tools using Gamification elements to enhance the benefits of social software. However, exemplary case studies show that current Gamification technology only supports rather quantitative than qualitative work improvement (see e.g. Amriani et al. 2013; Blohm and Leimeister 2013; De-Marcos et al. 2014; Farzan et al. 2008), hence not considering the above-described general criticism of Gamification such as regarding the overjustification effect. At the same time, the vendor market is on fluctuation and an overview is still missing. This article therefore evaluates leading social software like IBM Connections, Jive or Microsoft SharePoint, regarding the support of Gamification elements to shed light on the market. Besides the leading players third party vendors offering Gamification add-ons are considered for the market review. On this basis we will critically reflect how Gamification is used in social software.

The remainder of the paper is structured as follows. First in chapter 2 the related scientific work about social collaboration software and their adoption is provided. The objective of this chapter is to clarify why projects of social collaboration projects often suffer from poor acceptance among the employees. In this context, Gamification is introduced and described. Afterwards, the impact of using Gamification mechanisms to support social collaboration is reflected. Based on this, in chapter 3 a market review of Gamification software for leading social collaboration tools is presented to show which technical capabilities are currently available. Afterwards the Gamification functions are presented and critically discussed in chapter 4. The article ends with a conclusion, summarizing the key-findings of the critical review and providing advice for practitioners.

## 2 Literature Review and Theoretical Background

### 2.1 Social Collaboration

Literature reveals that collaboration describes the efforts of multiple individuals towards a mutually desired outcome (Briggs et al. 2006). This means that more than two people work together to achieve a common goal. "Collaboration is a special type of process that includes communication, coordination and cooperation" (Fan et al. 2012). In addition to this it is powerful for solving problems, making decisions and building consensus (Straus 2002). Additionally, there is no need to be physically present at the same place and time. Due to geographic or temporal reasons more and more people collaborate with each other via virtual technologies (Fan et al. 2012). One important aspect for providing a high level of collaboration performance is an adequate collaboration tool.

There are several theories trying to designate a set of principles how to select the optimal tool for achieving the most successful collaboration process (Fan et al. 2012). Task-technology fit theory (Zigurs et al. 1999) and Media Richness Theory (Daft and Lengel 1986) assert that the medium used for team communication needs to be well adapted to the type of information. Process virtualization theory discusses the suitability of different processes to be conducted virtually. There are four cases which are less appropriate for virtualization: human sensory experience, social context, time control and identity control (Fan et al. 2012). Schubert and Williams (2013) pointed out that one of the innovations of recent years was the appliance of the attribute "social" to the workplace. Companies increasingly pick up the concept of social media platforms like Facebook or Twitter and offer collaboraton technology, which support employees' interaction and exchange of employee-generated content across the whole enterprise, possibly affecting formal aspects of the organization including hierarchies and processes (Riemer et al. 2015; Stieglitz et al. 2014b; Bögel et al. 2014)

Collaboration technologies in general are "computer-based applications that support selected groups or specialized teams that work in various industries to develop new knowledge" (Lamb and Dembla 2013, p. 94). Sarrel (2010) states, that document-centric traditional collaboration tools are not sufficient to drive innovation and productivity. It is rather important to be able to leverage voice, video, presence information and instant messaging. User profiles are also key components of social software. Workers can build their personal brand by creating their own profile, share content and experiences, find expertise and offer their own knowledge (Sarrel 2010). According to the McKinsey Global Institute report two-thirds of the estimated economic value is due to improved communication and collaboration but a lot of companies are still missing a potentially "huge prize". Over 900 billion USD in annual value could be unlocked by products and services that facilitate social digital interactions. This is why enhancing the adoption of social collaboration technologies is an important process to manage.

One of the most highly cited models is the Technology Acceptance Model, which presents a basis to measure the impact of external factors on internal beliefs, intentions and attitudes concerning user adoption of Information Systems (Davis 1985). Due to this model external variables like pre-existing familiarity with social media can be useful so that users do not have to learn specific designs and applications within enterprise social software from scratch. These individual factors positively influence the perceived usefulness and ease of use. Other important aspects are the task complexity, organizational culture of the company and knowledge strategy. In the context of social software collaboration can only proceed when the participants have the necessary trust into achieving the goal through the new system (Lawson et al. 2007). de Oliveira and Watson-Manheim (2013) assert that the adoption and frequent usage of social software is not a controlled process but underlies a dynamic process. Old and new processes can affect the adoption of social media tools, which may be "constrained by existing processes but may also trigger creation of new ones" (de Oliveira and Watson-Manheim 2013, p. 2).

### 2.2 Gamification

An often cited definition in literature describes Gamification „as the use of game elements and techniques in nongame contexts" (Deterding et al. 2011, p. 2). From the market service perspective Gamification can be seen as „the process of enhancing a service with affordances for gameful experiences in order to support user's overall value creation" (Huotari and Hamari 2012, p. 19). Zichermann and Cunningham (2011) consider Gamification to be a process of game thinking that motivates users to perform particular tasks to solve problems or engage with customers. According to Shang and Lin (2013) games can be a powerful way to influence and change behavior in any setting.

According to Gabe Zichermann, CEO of Gamification Co, the early adoptions of Gamification occurred in response to employee dissatisfaction leading to disengagement. A recent Gallup poll reveals that two-thirds of the US workforce are disengaged or unengaged (Burmeister 2014). Yet, engagement of employees can create a 240% increase of performance related outcomes. Especially for businesses that are facing generational workforce shifts, Gamification could be useful (Burmeister 2014). Also Lounis et al. (2014) found out that participants experience more fun if they can collaborate with others towards a common goal. Another positive aspect of collaboration in gamified IS is the effect of 'social facilitation' which occurs when groups achieve better results than individuals (Zajonc 1965). Peischl et al. (2014) stated that "gamification acts as a layer on top of social collaboration software" and Rampoldi-Hnilo and Snyder (2013) even consider mobile workers to be the perfect audience for gamified applications. This reveals that the phenomenon of Gamification is more and more to be integrated in Information Systems in the business context. However, the above-cited literature does not distinguish between incentive mechanisms to improve the quantity or quality of work, relationships or others.

Gamification includes several game design elements like points, badges, leaderboards, rewards, levels, quests, challenges and virtual loops amongst others (Domínguez et al. 2013; De Paoli et al. 2012; Zichermann and Cunningham 2011). Those need to be implemented in the process of the transformation that incorporates game elements in the selected context. The motivation in using gamified elements lies in the satisfaction of fundamental human needs and desires, as the desire for reward, self- expression, altruism or competition (Bunchball Inc. 2010). In addition to this, the adequate combination of game mechanisms and dynamics shall create a motivating, emotional and entertaining interaction (Neeli 2012). In this context, suitable systems have the potential to set the user into a state of „flow" (Csikszentmihalyi 1991), where the user experiences a state of deep concentration. One of the conditions, which have to be fulfilled for a person to reach a state of flow, is an adequate balance between challenge and skill. Therefore, in Gamification environments, it is important for a task to match the users' skill level, where the user is neither under-challenged nor over-challenged (Groh 2012). Furthermore gamified applications have to offer tasks in an interesting way, handing out "juicy" feedback (Groh 2012).

These motivating processes can be useful for adapting and using new or existing Information Systems that otherwise often fail to meet their goals (Hsieh and Wang 2007). Especially intrinsic factors are important for motivating a certain behavior (Deci and Ryan 2000). Intrinsic motivation means the process of doing something due to satisfaction from the activity itself while extrinsic motivation, in contrast, implies an activity due to the prospect of an external outcome (Deci and Ryan 2000). Shauchenka et al. (2014) pointed out, that rewarding the quantitative performance of a user lead to a shift in motivation, where the user no longer enjoys the work itself but instead focuses on gaining points. According to the Goal Contents Theory (GCT) (Vansteenkiste et al. 2006) of Self Determination Theory (SDT) game elements that include monetary oriented goals can be seen as extrinsic goals whereas achievements to learn or improve in a certain activity lead to intrinsic motivation. While especially elements like points, badges, leaderboards and levels are used in the context of Gamification design, it seems that those elements are not adequate to make Gamification successful (Chorney 2012). While those elements may increase the performance of participating users (Mekler et al. 2013) it has been shown, that the removal of those elements could interrupt user interaction on the provided system (Amriani et al. 2013). A reason for this might be, that extrinsic rewards, punishments or regulations could diminish intrinsic motivations when individuals start to see the reward as the actual reason for performing an activity instead of doing so for their own interest or enjoyment (DeCharms 1968). This effect could be demonstrated in a simple experiment of Kohn 1999), where he could show, that children getting paid for drawing pictures, produced more pictures, but in lesser quality. After the payment was interrupted, the children did not draw as much as they did before. In a situation, where the quality of content is no longer taken into account, users' interest might shift and as a consequence, the users are may not longer be interested in contributing qualitative content (Shauchenka et al. 2014). Therefore the motivation shifts from intrinsic to extrinsic motivation and the user may get more motivated by gaining points, than by generating qualitative work (Shauchenka et al. 2014).

# 3　Market Review

## 3.1　Methodology

There are about 100 social software vendors on the market (Mladjov 2013). Considering all of them for this review would burst the scope of this article. Thus, first the number of social software vendors for the evaluation sample needs to be set. Therefore the latest Gartner report of social software in the workplace was used as a basis.[1] The analysts separate the market in four categories, (1) niche players, (2) visionaries, (3) challengers, and (4) leaders (Drakos et al. 2014). However, serveral approaches to differentiate the market in a first step exist and other business analyst may build on different samples. For the overriding goal of this paper - to conduct a critical market overview of gamification plug-ins for social software – the differentiation of Gartner is used. Gartner Inc is a well established and realiable market research institute. Neverthesless their segmentation underlies some restrictions, as they only consider vendors whoe are active on at least three continents and have a turnover above $ 50Mio. Hence, the goal of this article is to provide an overview of Gamification functions from established social software and from third party vendors. Particularly the third party vendors merely concentrate their activities on leading platforms with high market share. Following Gartner these are software vendors, *"which have established their leadership through early recognition of users' needs, continuous innovation, significant market presence, and success in delivering user-friendly and solution focused suites with broad capabilities"* (Drakos et al. 2014). Based on the classification of Gartner, the five "leading" vendors (IBM, Microsoft, Jive, Salesforce, Tibco Software) build the unit for this analysis. The market examination was conducted in late 2014. Meanwhile Salesforce has been acquired by Microsoft in 2015.

Next the products of the vendors were selected. IBM, Jive and Tibco offer one social software product. In that case this one was chosen for the market review. Microsoft and Salesforce offer more than one social software product. Microsoft offer SharePoint and Yammer as their social software products. Microsoft intends to concentrate their social activities on Yammer, but right now SharePoint is the leading product. Hence both products were considered in the market review. Whereas Salesforce clearly distinct their products Chatter and Communities. Communities concentrates on the support to establish larger-scale communities of partners and customers, whereas Chatter is a social networking tool for employee networking supporting collaboration features (Drakos et al. 2014). Due to the missing focus on collaboration, Communities was skipped and Chatter was selected for the sample.

In a next step the product websites of the software vendors have been independently evaluated by two different researchers, regarding the availability of Gamification functions. By doing so several third party vendors offering plug-ins or add-ons for leading social software could be identified. However, not every social software vendor offered the requested information on their website. Thus, a keyword based web search was conducted. Table 1 shows the applied search strings used for the investigation via google. The keywords were validated by a first pre-test and adjusted based on the first results during the search process. From the results page the first 10 results (1st page) were analyzed, as these results represent the most relevant and important search results and only 7% of the users click to second results page.[2] Moreover we had to limit the number of pages to view. The method was used to gain first insights in the market.. This approach resulted in an overview (Table 3) providing a short description and available Gamification features (e.g. highscores, badges, and quests).

| Social software product (vendor) | Search strings to identify third party vendors (plug-ins and add-ons) |
|---|---|
| Chatter (Salesforce) | „chatter gamification"; „chatter gamification plugin"; „chatter gamification enterprise "chatter gamification standard |
| IBM Connections (IBM) | „ibm connections gamification"; „ibm connections gamification plugin"; „ibm connections enterprise-gamification"; "ibm connections gamification standard |
| Sharepoint (Microsoft) | „sharepoint gamification"; „sharepoint gamification plugin"; |

---

[1] https://www.gartner.com/doc/2836617/magic-quadrant-social-software-workplace
[2] http://www.advancedwebranking.com/google-ctr-study-2014.html

| Social software product | Third party vendor |
|---|---|
| | „sharepoint gamification-enterprise "sharepoint gamification standard |
| tibbr(Tibco Software) | „tibbr gamification"; „tibbr gamification plugin"; „ tibbr enterprise-gamification"; ";; „tibbr gamification standard"; |
| JIVE (Jive) | „jive gamification"; „jive gamification plugin"; „jive gamification-enterprise"; "jive gamification standard" |
| Yammer (Microsoft) | „yammer gamification"; „yammer gamification plugin"; „yammer enterprise-gamification"; „yammer gamification standard |

*Table 1. Search strings to identify third party vendors.*

## 3.2 Results

Based on the search results derived by the keywords shown in table 1, the two indepented researchers were able to identify the third party vendors (plug-ins / add-ons) shown in Table 2. Both visited all websites generated by the google search and manually created the vendor list. These third party vendors are selected for the market review regarding Gamification features. The following sub-chapters aim to shortly present the products and their main Gamification functions.

| **Social software product** | **Third party vendor** |
|---|---|
| Chatter (Salesforce) | • Chatter Answers<br>• The Chatter Game<br>• RedCritter |
| IBM Connections (IBM) | • Kudos Badges<br>• Badgeville<br>• Nitro Bunchball |
| Sharepoint (Microsoft) | • Badgeville<br>• Beezy<br>• RedCritter<br>• Attini |
| tibbr (Tibco Software) | • (Announced Partnership with) Badgeville |
| JIVE (JIVE) | • Badgeville<br>• Nitro Bunchball |
| Yammer (Microsoft) | • Face Game<br>• RedCritter<br>• Badgeville |

*Table 2. Overview of selected social software products and third party vendors*

### 3.2.1 Chatter

Salesforce Chatter does not contain any Gamification functions, yet three third party plug-ins could be identified (Chatter answers, the Chatter game and RedCritter), which cover this domain. Using Chatter answers, the users gain points for certain activities within the network, particularly for answers. The peculiarity within this tool is a live board, which always displays which user is the most active within the community just at the moment.

The Chatter game consists of posts that can be commented by and within the community. This feedback affects the score, which is awarded for the post. The score is displayed, as well as the given feedback, in the user's personal profile. The prerequisite for this tool is an already active community. Also, this tool fosters an active community. When using RedCritter for Chatter, the users are engaged to earn rewards for activities, which they can share among the community members. This allows a competition between the employees (see 3.2.3).

### 3.2.2　IBM Connections

Similar to Chatter from Salesforce, IBM does not offer any game mechanism within their collaboration software IBM Connections by default. To cover this domain, plug-ins are necessary. One plug-in is Kudus, which bases on reaching levels in different areas. In those different areas points are gained via activities so that new levels between "Newbie" and "Hall of Fame" might be reached. For each new level the employee gets a new badge, shown in the user's profile. Another feature is the "thanks-function", which allows the user to give thanks to others for dispatching a task. This direct feedback by the co-workers motivates the users because they get a reward for their work. Another aspect is that the users can see how reliable another user is before they give him a task.

Another available plug-in is Nitro Bunchball. This plug-in allows the administrator to create different missions, which have to be accomplished by the users so that points and badges can be gained. Each mission is individually adjustable and there is the possibility of creating different blocks out of several missions. Furthermore the plug-in Badegeville allows the users to earn points and badges in two ways: either by active usage (sharing, participating in discussions) of the social-collaboration-tool (IBM Connections) or by fulfilling given missions, which contribute to a better acknowledgement of the tool. The level and badges, which are achieved by points are shown in the user's profile and in the leaderboard.

### 3.2.3　SharePoint

SharePoint is the only social software product that offers Gamification features by default. Since version 13.2 SharePoint contains two Gamification features: first, the community template, which offers a discussion's list based on various sites that are available for discussing. For each post points are gained and from a certain score on badges for one's personal profile are achieved. Second, SharePoint provides an e-learning-feature. The users are gaining points as well as awards, which are shown in one's personal profile if they complete a so called class. E-learning classes can be online courses or classes, in which users can learn alone or together with a teacher and other employees. Besides these SharePoint's Gamification features there are several third party plug-ins available. First, Badgeville offers the same features for SharePoint as for IBM Connections (see 3.2.2). The second one is RedCritter, which is also available for Chatter and Yammer. In sum, the employees get rewards in terms of points or badges for participation and further education. Those help the users to gain a higher level and can be reached via activities, both online as well as those offline (e. g. via QR code). Both level and badges are saved in the user's profile and can be seen by other staff in a leaderboard, which can be searched for special levels and badges by the team leaders.

Another plug-in by Beezy does not automatically rewards the users, but the users mutually reward themselves. Beezy also offers the possibility of giving feedback, which is presented in one's personal profile. The last identified plug-in for SharePoint is Attini. Attini allows to gain badges by certain social activities within SharePoint, which can be shown in one's personal *profile*.

### 3.2.4　tibbr

For Tibco's tibbr neither included Gamification features nor third party vendors offering Gamification plug-ins or add-ons could be identified. However in 2012 Tibco announced a partnership with Badgeville. It was intended to make it possible for tibbr users to earn contextually relevant rewards mapped to their expertise and contributions within the tibbr platform. So far it was not possible to identify any plug-in on the Badgeville and tibbr webpage.

### 3.2.5　JIVE

The social software Jive does not contain any Gamification functions yet. However two third party plug-ins could be identified, enhancing the functionalities of Jive. The first one is Badgegeville and the second one is Nitro Bunchball. Both plug-ins are also available for IBM Connections and offer the same Gamification features for both social software products. For more information please see chapter 3.2.2.

### 3.2.6　Yammer

As the other social software products, except SharePoint, Yammer does not cover any Gamification features. Anyway there are third party plug-ins available, such as face game, a game in which the other staffs' faces are shown and have to be recognized by the player. The users gain points for correct answers and are able to compare each other within a ranking. The idea behind this is that the team gets to know each other and the workers know who their teammates are, what can reduce the lack of communication and improve the working atmosphere. The tool is suitable to integrate new members into a team or building up a new team. At least the already known plug-ins RedCritter and Badegeville are available for Yammer as well with the Gamification mechanisms described in 3.2.3 and 3.2.2.

## 4　Discussion

Summarizing the market review (see Table 3) it can be stated that especially third party plug-ins are prominent on the market for Gamification elements in social collaboration software. Only Microsoft SharePoint offers Gamification features by default. One possible reason for this might be, that due to the high quality of those plug-ins the vendors of social collaboration tools may quit in-house development of Gamification elements and cooperate with third party suppliers. According to the attributes and purposes of the plug-ins they could be divided into two categories.

The first category consists of tools, which aim at the encouragement of the staff to improve their education. For accomplishing further development points and rewards are provided, which shall portray a particular know-how of the user. Tools within this category are Kudos Badges, Face Game, RedCritter, and E-Learning-Function (SharePoint). The second category aims at the motivation of the staff to enhance their interaction within social collaboration software as well as motivating the employees to use the adapted tools actively and regularly. Therefore, this category does not aim to upgrade the staff's education but on the implementation of social collaboration tools in the workflow. The main aim is to create and motivate a community through social software. Solutions within this category are Badgeville, Attini, Nitro Bunchball, Chatter Answer, Chatter Game, and Community Template (SharePoint). Moreover it could be observed that the third party vendors try to offer their services for a multitude of social software products. Especially Badgeville and RedCritter follow that approach. They put themselves in the position of Gamification specialists for social software products and collaboration. Most of the offered Gamification mechanisms are leaderboards, badges and points, less often implemented are challenges, quest, levels and rewards. Ownerships, bonus or status are not used in our sample.

| Social Software | Third Party Vendors | Badges | Challenges/Quests | Community Collab. | Leaderboards | Levels | Points | Rewards/Feedback |
|---|---|---|---|---|---|---|---|---|
| Chatter | Chatter Answers | | | | x | | x | |
| | The Chatter Game | | | | x | | x | |
| | RedCritter | x | | | x | x | x | x |
| IBM Connections | Kudos Badges | x | | x | x | | x | |
| | Badgeville | x | x | | x | x | x | |
| | Nitro Bunchball | x | x | | x | x | x | |
| SharePoint | Badgeville | x | x | | x | x | x | |
| | Beezy | x | | | | | | x |
| | RedCritter | x | | | x | x | x | x |
| | Attini | x | | | | | | |
| SP: Community Template | | x | | | x | | x | |

| | | | | | | | |
|---|---|---|---|---|---|---|---|
| SP: E-Learning Feature | | | | | x | x | |
| tibbr | Not available | | | | | | |
| Jive | Badgeville | x | x | | x | x | x |
| | Nitro Bunchball | x | x | | x | x | x |
| Yammer | Face Game | | | | x | | x |
| | RedCritter | x | | | x | x | x | x |
| | Badgeville | x | x | | x | x | x |

*Table 3. Overview of applied game mechanisms*

While most of the third party vendors focus on reward mechanisms, none of the plug-ins taken into account did measure or reward qualitative performance. Most of the Gamification plug-ins had a strong focus on rewarding quantitative user performance. Since points, badges and leaderboards have proven their influence in Gamification systems by increasing users' participation and communication activity, the trend of adding those elements in social collaboration software seem to be legitimate. However, the „efficiency" is quite questionable, since studies of Amriani et al. (2013) and Kohn (1999) could prove the risk of an upcoming overjustification effect, where users lose the intrinsic motivation in their work. Especially the one-sided view on rewarding quantitative performance is quite alarming, what studies of Schubert et al. (2014) and Shauchenka et al. (2014) also criticize. Therefore the long term effectiveness of such systems has to be questioned in general. It seems that the trend of current Gamification implementations mainly focuses on increasing quantitative performance in the first IT-adoption phase and that long-term goals are missing. The goal of setting a users into a state of „flow" has been ignored in all implementations. According to literature a state of flow can only be reached if the users' tasks match the users' skills (Csikszentmihalyi 1991). Due to the description of Groh (2012) tasks therefore need to provide interesting challenges, which we were not able to see in any of the observed Gamification features or reported case studies.

## 5   Conclusion

The benefit of collaboration software in organizational environments as well as the tremendous diffusion of tools has been well documented in prior work. Organizations are faced to manage the adoption of these collaborative tools and therefore need to increase the acceptance and usage motivation. One way to solve this problem can be Gamification. As prior work showed, Gamification elements in social software are able to improve users' engagement through extrinsic motivation and therefore lead to a better acceptance of the system. However, while Gamification has been adequately discussed in literature, the market for collaborative software is still young and fluctuating. Research has ignored several market determining tools as well as third party vendor solutions. Consequently an overview of major collaboration tools and the possibilities to use the tools in a gamified way was missing. In this work we tried to fill this gap by analysing major social software solutions as well as third party vendors offering Gamification add-ons to those. We found out, that while social software tools primarily do not include Gamification elements, plenty of third party vendors offer add-ons to embedded Gamification elements to the software. In addition we make several research contributions and highlight that current Gamification approaches still ignored warnings of scientists: although the simple usage of extrinsic motivators holds the risk of lowering users' intrinsic motivation and causing the denial of the system or even the work task in the long run, Gamification features still focus on exactly those mechanisms. Most notably, promoting intrinsic motivation has been mainly neglected in the implementations. The aim in current implementations is to increase the quantitative performance instead of the qualitative performance. Since the user gets rewarded for doing nothing but quantitative work, according to Groh (2012) a state of flow can't be reached. Overall, our findings show a gap between current research and practical usage.

As any research, this article comes with some limitations. This research is still in progress. Although our work focuses on collaboration software with high market shares, we were only able to analyze a small group of tools and third party vendors. In any case, the issue of different Gamification types deserve future research attention in the context of Gamification of collaborative environments. Implementations also need to support the emergence of intrinsic motivation. Research should support the development by identifying new strategies and Gamification elements providing corresponding needs.